\documentclass{article}
\usepackage{spconf,amsmath,graphicx}
\usepackage{epstopdf,booktabs,multicol,multirow,enumitem,footnote,type1cm,amssymb,calc}
\usepackage{color,xcolor}
\usepackage[linesnumbered,ruled,vlined]{algorithm2e}
\SetKwInput{KwInit}{Initialization}                
\SetKwInput{KwOutput}{Output}              

\SetCommentSty{mycommfont}

\title{TrimTail: Low-Latency Streaming ASR with simple but effective spectrogram-level length penalty\vspace{-10pt} }
\name{
    \begin{tabular}{c}
        Xingchen Song$^{1,2,3\dagger}$\thanks{$\dagger$ Equal Contribution and Corresponding Author.},~Di Wu$^{2,3,\dagger}$,~Zhiyong Wu$^1$,~Binbin Zhang$^{2,3}$,~Yuekai Zhang$^{3}$\\Zhendong Peng$^{2,3}$,~Wenpeng Li$^{2}$,~Fuping Pan$^{2}$,~Changbao Zhu$^{2}$
    \end{tabular}
    \vspace{-10pt}
}
\address{
    $^1$Tsinghua Univ., Beijing, China~~ 
    $^2$Horizon Inc., Beijing, China~~
    $^3$WeNet Open Source Community \\
    xingchen.song@horizon.ai$^{\dagger}$,~~di02.wu@horizon.ai$^{\dagger}$
    \vspace{-20pt}
}

\begin{document}

\maketitle

\begin{abstract}
In this paper, we present \textit{TrimTail}, a simple but effective emission regularization method to improve the latency of streaming ASR models. The core idea of \textit{TrimTail} is to apply length penalty~(i.e., by trimming trailing frames, see Fig.~\ref{fig:spectrim}-(b)) directly on the spectrogram of input utterances, which does not require any alignment. We demonstrate that \textbf{\textit{TrimTail is 
computationally cheap and can be applied online and optimized with any training loss or any model architecture on any dataset without any extra effort}}
by applying it on various end-to-end streaming ASR networks either trained with 
 CTC loss~\cite{icml06/ctc} or Transducer loss~\cite{DBLP:journals/corr/abs-1211-3711}.
We achieve 100 $\sim$ 200ms latency reduction with equal or even better accuracy on both Aishell-1 and Librispeech. Moreover, by using \textit{TrimTail}, we can achieve a 400ms algorithmic improvement of User Sensitive Delay (USD) with an accuracy loss of less than 0.2.
\end{abstract}


\section{Introduction}
\label{sec:intro}

End-to-end (E2E) models such as CTC~\cite{icml06/ctc}, recurrent neural network transducer (RNN-T)~\cite{DBLP:journals/corr/abs-1211-3711}, and attention-based encoder-decoder (AED)~\cite{ic16/LAS,ic18/speechtransformer} gained more and more attention over the last few years. Compared with the conventional hybrid automatic speech recognition (ASR) framework, E2E models not only extremely simplified training and decoding procedure but also show superior performance in the measure of word error rate (WER). So applying the E2E models to real-world productions becomes necessary. However, deploying E2E systems is not trivial and there are a lot of practical problems to be solved. For example, apart from being accurate and compact, ASR systems need to decode speech in a \textit{streamable} fashion with low user-perceived latency~\cite{shangguan21_interspeech}.
In this work, we mainly focus on reducing streaming recognition latency to achieve a low-latency user experience when integrating ASR into other applications like Real-Time Video Subtitles and Smart Home Assistants.



Streaming ASR models aim to emit each hypothesized word as quickly and accurately as possible. However, emitting fast without degrading quality, as measured by WER, is highly challenging~\cite{yu2021fastemit}.
Existing approaches aiming to improve streaming latency heavily rely on loss function design and force alignments. For example, Li \textit{et al.}~\cite{DBLP:conf/icassp/LiCSPHSW20} apply Early and Late Penalties to training loss to enforce the prediction of eos (end of sentence) within a reasonable time window.
Others use Constrained Alignments~\cite{DBLP:conf/interspeech/SakSRB15} to extend the penalty terms to each word.
While being successful in reducing delay, these approaches suffer from significant accuracy regression and also require additional token alignment information. To avoid any requirement of token alignments, Yu \textit{et al.}~\cite{yu2021fastemit} propose \textit{FastEmit} to directly modify the forward-backward per-sequence probability for Transducer loss~\cite{DBLP:journals/corr/abs-1211-3711} to encourage faster token emission. Recently, Tian \textit{et al.}~\cite{DBLP:journals/corr/abs-2210-07499} apply bayes risk function to the forward-backward algorithm of CTC to control CTC alignment with better WER-latency trade-off.

\begin{figure}[!htp]
    \centering
    \includegraphics[scale=0.22]{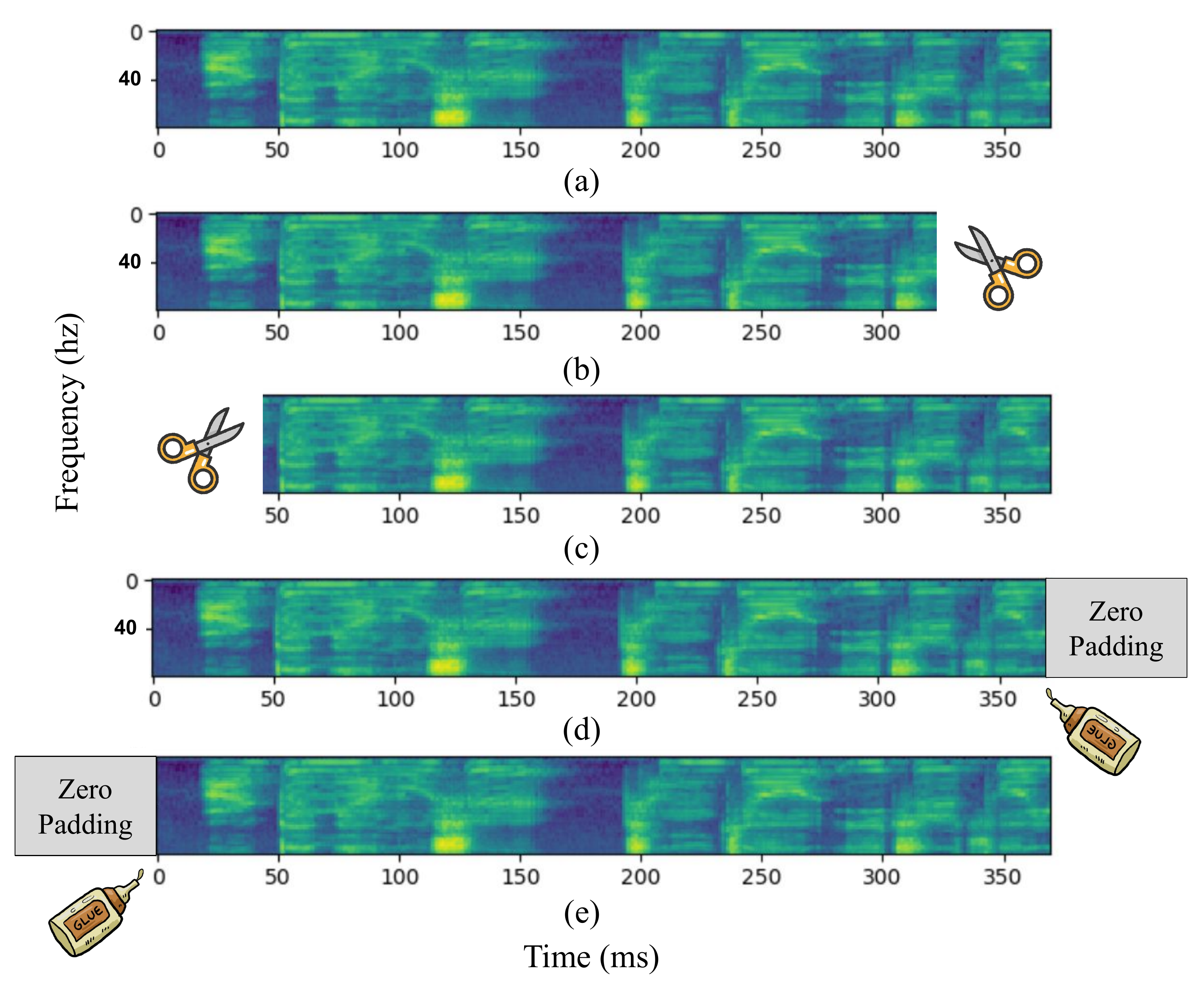}
    \vspace{-10pt}
    \caption{Length penalty applied to the base input. From top to bottom: (a) Spectrogram of the base input with no penalty, (b) Trim trailing frames~(\textit{TrimTail}), (c) Trim leading frames (\textit{TrimHead}), (d) Pad trailing frames (\textit{PadTail}) and (e) Pad leading frames (\textit{PadHead}).}
    \label{fig:spectrim}
\end{figure}

Unfortunately, all those methods are bound to specific loss functions (Transducer~\cite{DBLP:conf/icassp/LiCSPHSW20,DBLP:conf/interspeech/SakSRB15,yu2021fastemit} or CTC~\cite{DBLP:journals/corr/abs-2210-07499}) and thus cannot achieve a \textbf{Universal Solution} towards reducing streaming latency on any ASR model without any pain.
In this work, we propose a novel emission regularization method for all kinds of ASR models, which we call \textit{TrimTail}. \textit{TrimTail} is a kind of length penalty designed to be directly applied on the spectrogram of input utterances, rather than modify loss calculation. Empirically, we argue that by trimming trailing frames~(Fig.~\ref{fig:spectrim}-(b)):
\vspace{-5pt}
\begin{itemize}
    \item The speech-text alignment space will be \textbf{``squeezed''} and trailing tokens are forced to align with preceding speech frames, pushing forward the emission of previous tokens to meet the monotonic requirement of alignment, thus leading to significantly lower latency.
    \vspace{-5pt}
    \item Models are encouraged to predict the trailing tokens even before they were spoken since corresponding frames were trimmed during training.
    \vspace{-5pt}
\end{itemize}
\vspace{-2pt}

To validate our conjectures, we also propose to apply other length penalty policies opposite to \textit{TrimTail}, i.e., trimming leading frames (Fig.~\ref{fig:spectrim}-(c)) to delay the emission of leading tokens and its subsequent counterparts or padding zero frames at the ending~(Fig.~\ref{fig:spectrim}-(d)) or beginning~(Fig.~\ref{fig:spectrim}-(e)) of the input spectrogram to \textbf{``expand''} the alignment space. Detailed analysis can be found in section~\ref{sec:expablation}.

\textit{TrimTail} has many advantages over other regularization methods to reduce emission latency in end-to-end streaming ASR models: 
(1) \textit{TrimTail} does not require any speech-text alignment information.
(2) \textit{TrimTail} is easy to plugin any ASR model (either trained by Transducer or CTC) on any dataset without any extra effort. (3) There is no additional training, serving, or engineering cost to apply \textit{TrimTail}. Instead, one can simply implement such a strategy and benefit from length trimming to achieve even more computationally efficient training.

We apply \textit{TrimTail} on various end-to-end streaming ASR networks and training losses
and achieve 100 $\sim$ 200ms latency reduction with equal or even better accuracy on both Aishell-1 and Librispeech. Moreover, by using \textit{TrimTail}, we can achieve a 400ms algorithmic improvement of User Sensitive Delay (described in section~\ref{sec:metrics}) with an accuracy loss of less than 0.2.

\vspace{-8pt}

\vspace{-20pt}
\section{Methodology}
\vspace{-5pt}

In this section, we first delve into All-in-One U2++, an enhanced version of U2++~\cite{DBLP:journals/corr/abs-2106-05642}, which not only unifies streaming and non-streaming two-pass bidirectional ASR, but also integrates CTC, Transducer, and AED into a unified architecture. Then we introduce \textit{TrimTail} as a length penalty method to regularize the emission latency.

\vspace{-10pt}
\subsection{All-in-One U2++}
\vspace{-10pt}

\begin{figure}[!h]
    \centering
    \includegraphics[scale=0.3]{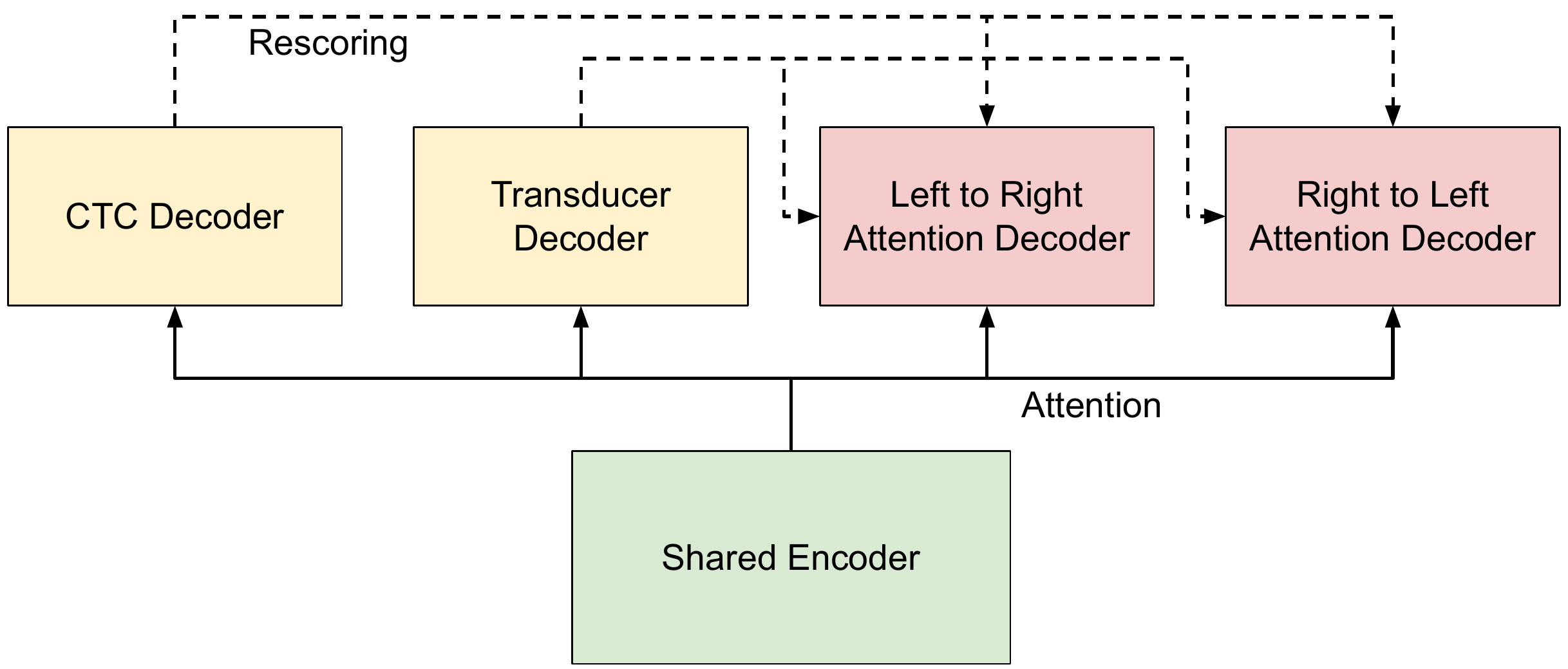}
    \caption{2-pass CTC, Transducer and AED joint architecture.}
    \label{fig:u2++}
\end{figure}
\vspace{-10pt}

The model architecture used in this paper is shown in Figure~\ref{fig:u2++}. It contains five parts, a Shared Encoder that models the context of the acoustic features, two Decoders (CTC and Transducer) that models the alignment of the frames and tokens, a Left-to-Right Attention Decoder (L2R) that models the left tokens dependency, and a Right-to-Left Attention Decoder (R2L) model the right tokens dependency. The Shared Encoder consists of multiple Transformer~\cite{nips17/transformer} or Conformer~\cite{DBLP:conf/interspeech/GulatiQCPZYHWZW20} encoder layers.
Due to space limitations,
we would refer the reader to the individual source codes~\footnote{https://github.com/wenet-e2e/wenet\label{web}} for more details of each model architecture.


\vspace{-10pt}

\subsection{TrimTail}
\vspace{-15pt}
\begin{algorithm}[!ht]
    \label{algo:TrimTail}
    \DontPrintSemicolon
      \KwInit{$T_{max}$, $T_{min}=1$}
      \tcc{$data$ is 80-mel spectrogram}
      \For{data in batch} {
        sample $t \sim Uni(T_{min}, T_{max})$ \\
        $len \gets length(data)$ \\
        \If{$t < len / 2$} {
            $data \gets data(0, len - t)$
        }
      }
    \caption{TrimTail}
\end{algorithm}
\vspace{-15pt}

We aim to construct a length penalty policy that directly acts on the input sequences, which serves as the emission regularization strategy. As shown in Fig.~\ref{fig:spectrim}-(b) and Algorithm~\ref{algo:TrimTail}, \textbf{\textit{TrimTail is computationally cheap and can be applied online and optimized with any training loss or any model architecture on any dataset without any extra effort}}. The codes and configurations used in this paper have all been released~\footnote{https://github.com/wenet-e2e/wenet/pull/1487}.

\vspace{-10pt}
\section{Experiments}
\label{sec:exp}
\vspace{-10pt}

\subsection{Latency Metrics}
\label{sec:metrics}
\vspace{-5pt}

\begin{figure*}
    \centering
    \vspace{-10pt}
    \includegraphics[scale=0.5]{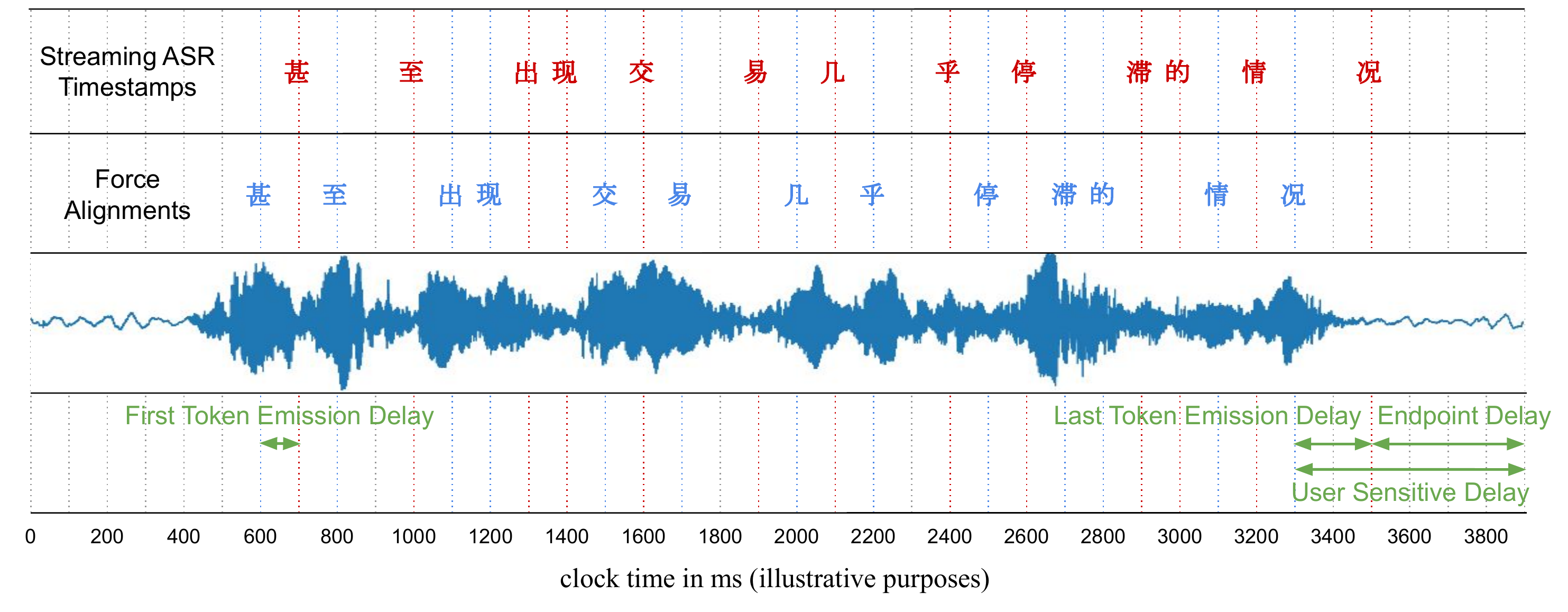}
    \vspace{-15pt}
    \caption{A visual illustration of timeline and latency metrics of a streaming ASR system.}
    \vspace{-12pt}
    \label{fig:latency}
\end{figure*}

Similar to \cite{yu2021fastemit,shangguan21_interspeech}, Our latency metrics of streaming ASR are motivated by real-world applications like Real-time Video Subtitles and Smart Home Assistants. In this work we mainly measure four types of latency metrics described below: (1) First Token emission Delay~(FTD), (2) Last Token emission Delay~(LTD), (3) Average Token emission Delay~(AvgTD), and (4) User Sensitive Delay~(USD). A visual example of some latency metrics is illustrated in Figure~\ref{fig:latency}. For all metrics, we report both 50-th (medium) and 90-th percentile values of all utterances in the test set to better characterize latency by excluding outlier utterances.

\textbf{First Token emission Dealy~(FTD)} (100ms, in this example) is defined as the timestamps difference of two events as illustrated in Figure~\ref{fig:latency}: (1) when the first token is emitted in the streaming recognition result, (2) the timestamp of the first token estimated by force alignments. Similar to FTD, \textbf{Last Token emission Delay~(LTD)} (200ms, in this example) is defined to measure the emission latency for the final token. FTD and LTD are especially descriptive of user experience in real-world streaming ASR applications like Video Subtitles.
To analyze emission delay for intermediate recognition result, we also report \textbf{Average Token emission Delay~(AvgTD)} (162ms, in this example) as described below:
\vspace{-15pt}
\begin{equation}
    AvgTD = \frac{\sum\limits_{k=1}^{K}{Timestamps_k} - \sum\limits_{k=1}^{K}{Alignments_k}}{K}
\end{equation}
\vspace{-15pt}

Where $K$ is the total number of tokens in the current sentence, $Timestamps$ and $Alignments$ are generated from streaming recognition results and force alignments, respectively. Although AvgTD does not correlate with user experience as strongly as FTD and LTD, it serves as a fair metric that represents an overall latency improvement across the whole utterance.

Improving token emission latency is essential for applications like Real-Time Video Subtitles. However, such metrics may not be suitable for Smart Home Assistants, since they do not account for interactions with other system components such as vad endpointers and subsequent intent understanding.


To better estimate the user experience for different applications, besides token emission latency, we also introduce the \textbf{User Sensitive Delay~(USD)} (600ms, in this example), a union of both LTD and Endpoint Delay (400ms, in this example), to measure the time between when the user finishes speaking and the decision is made to close the microphone. The endpoint model typically runs in parallel with the decoder and detects the end-of-query in order to close the microphone~\cite{shangguan21_interspeech}.
Since the accurate analysis of USD requires ground-truth human-annotated end-of-speech labels which are not provided in open-sourced datasets, we instead measure the algorithmic improvement of USD when comparing models trained with/without \textit{TrimTail} in section~\ref{sec:improve_usd}. 

\vspace{-8pt}
\subsection{Dataset, Training Details and Model Architectures}

\begin{table*}[!htp]
    \centering
    \vspace{-15pt}
    \caption{WER (\textit{1st-pass Greedy Search / 2nd-pass Rescore}) and latency comparison of different \textit{TrimTail($T_{max}$)} across different models or losses. The Transducer Decoder and CTC Decoder are excluded in Conformer (CTC) and Conformer (Transducer), respectively. All models are jointly trained with AED loss. For Librispeech dataset, due to the mismatched modeling units between force alignment system (phonemes) and E2E system (byte pair encodings), we treat baseline timestamps as ``fake'' force alignments and only report relative improvements when applying \textit{TrimTail}.}
    \scalebox{0.8}{
        \begin{tabular}{c|c|cccccc}
        \toprule[1.5pt]
            Models & WER (\%) & FTD50 (ms) & FTD90 (ms) & LTD50 (ms) & LTD90 (ms) & AvgTD50 (ms) & AvgTD90 (ms) \\
        \midrule
        \multicolumn{8}{c}{Aishell-1 (test) 1/4 subsample \& 640ms chunksize} \\
        \midrule
            Transformer (CTC) & 6.92 / 5.63 & 70 & 120 & 30 & 90 & 68 & 92\\ 
             ~~\textit{+TrimTail(50)}  & \textbf{6.78} / 5.65 & 90 \textcolor{red}{($\uparrow$20)} & 140 \textcolor{red}{($\uparrow$20)} & 30 \textcolor{green}{($\sim$)} & 110 \textcolor{red}{($\uparrow$20)} & 101 \textcolor{red}{($\uparrow$33)} & 142 \textcolor{red}{($\uparrow$50)} \\
             ~~\textit{+TrimTail(70)}  & 6.83 / \textbf{5.61} & 60 \textcolor{green}{($\downarrow$10)} & 120 \textcolor{green}{($\sim$)} & -50 \textcolor{green}{($\downarrow$80)} & 30 \textcolor{green}{($\downarrow$60)} & 40 \textcolor{green}{($\downarrow$28)} & 68 \textcolor{green}{($\downarrow$24)} \\
             ~~~\textit{+TrimTail(100)}  & 7.20 / 5.75 & 40 \textcolor{green}{($\downarrow$30)} & 100 \textcolor{green}{($\downarrow$20)} & -100 \textcolor{green}{($\downarrow$130)} & -50 \textcolor{green}{($\downarrow$140)} & -34 \textcolor{green}{($\downarrow$102)} & -13 \textcolor{green}{($\downarrow105$)} \\
        \midrule
            Conformer (CTC) & \textbf{5.81} / \textbf{5.05} & 230 & 280 & 210 & 250 & 235 & 260 \\ 
             ~~\textit{+TrimTail(50)}  & 5.83 / 5.09 & 220 \textcolor{green}{($\downarrow$10)} & 260 \textcolor{green}{($\downarrow$20)} & 60 \textcolor{green}{($\downarrow$150)} & 120 \textcolor{green}{($\downarrow$130)} & 188 \textcolor{green}{($\downarrow$47)} & 210 \textcolor{green}{($\downarrow$50)} \\
             ~~\textit{+TrimTail(70)}  & 6.03 / 5.17 & 150 \textcolor{green}{($\downarrow$80)} & 200 \textcolor{green}{($\downarrow$80)} & -10 \textcolor{green}{($\downarrow$220)} & 50 \textcolor{green}{($\downarrow$200)} & 109 \textcolor{green}{($\downarrow$126)} & 132 \textcolor{green}{($\downarrow$128)} \\
             ~~~\textit{+TrimTail(100)}  & 6.48 / 5.41 & 100 \textcolor{green}{($\downarrow$130)} & 140 \textcolor{green}{($\downarrow$140)} & -80 \textcolor{green}{($\downarrow$290)} & -20 \textcolor{green}{($\downarrow$270)} & 26 \textcolor{green}{($\downarrow$209)} & 52 \textcolor{green}{($\downarrow$208)} \\
        \midrule
            Conformer (Transducer) & 6.96 / N/A  & 290 & 350 & 200 & 230 & 270 & 284 \\
            ~~\textit{+TrimTail(50)} & 7.09 / N/A  & 190 
            \textcolor{green}{($\downarrow$100)} & 250 
            \textcolor{green}{($\downarrow$100)} & 40 
            \textcolor{green}{($\downarrow$160)} & 80 
            \textcolor{green}{($\downarrow$150)} & 156 
            \textcolor{green}{($\downarrow$114)} & 175 
            \textcolor{green}{($\downarrow$109)} \\
            ~~\textit{+TrimTail(70)} & \textbf{6.94} / N/A  & 150 
            \textcolor{green}{($\downarrow$140)} & 210 
            \textcolor{green}{($\downarrow$140)} & -30 
            \textcolor{green}{($\downarrow$230)} & 20 
            \textcolor{green}{($\downarrow$210)} & 98 
            \textcolor{green}{($\downarrow$172)} & 121 
            \textcolor{green}{($\downarrow$163)} \\
            ~~~\textit{+TrimTail(100)} & 7.36 / N/A  & 130 
            \textcolor{green}{($\downarrow$160)} & 190 
            \textcolor{green}{($\downarrow$160)} & -60 
            \textcolor{green}{($\downarrow$260)} & -10 
            \textcolor{green}{($\downarrow$240)} & 58 
            \textcolor{green}{($\downarrow$212)} & 80 
            \textcolor{green}{($\downarrow$204)} \\
            \midrule
            Conformer (CTC + Transducer) & \textbf{6.34}/\textbf{5.69}  & 250 & 310 & 140 & 190 & 234 & 247 \\
            ~~\textit{+TrimTail(50)} & 6.47/5.78  & 180 
            \textcolor{green}{($\downarrow$70)} & 220 
            \textcolor{green}{($\downarrow$90)} & 10 
            \textcolor{green}{($\downarrow$130)} & 60 
            \textcolor{green}{($\downarrow$130)} & 138 
            \textcolor{green}{($\downarrow$96)} & 155 
            \textcolor{green}{($\downarrow$92)} \\
            ~~\textit{+TrimTail(70)} & 6.71/5.87  & 120 
            \textcolor{green}{($\downarrow$130)} & 180
            \textcolor{green}{($\downarrow$130)} & -80
            \textcolor{green}{($\downarrow$220)} & -20 
            \textcolor{green}{($\downarrow$210)} & 56
            \textcolor{green}{($\downarrow$178)} & 73
            \textcolor{green}{($\downarrow$174)} \\
            ~~~\textit{+TrimTail(100)} & 7.28/6.16  & 100 
            \textcolor{green}{($\downarrow$150)} & 170 
            \textcolor{green}{($\downarrow$140)} & -120 
            \textcolor{green}{($\downarrow$260)} & -90
            \textcolor{green}{($\downarrow$280)} & 5 
            \textcolor{green}{($\downarrow$229)} & 28
            \textcolor{green}{($\downarrow$219)} \\
        \midrule
        \multicolumn{8}{c}{Aishell (test) 1/4 subsample \& 320ms chunksize} \\
        \midrule
            Conformer (CTC) & \textbf{6.13} / \textbf{5.27} & 240 & 300 & 210 & 250 & 237 & 263 \\ 
             ~~\textit{+TrimTail(50)}  & 6.20 / \textbf{5.27} & 230 \textcolor{green}{($\downarrow$10)} & 280 \textcolor{green}{($\downarrow$20)} & 50 \textcolor{green}{($\downarrow$160)} & 120 \textcolor{green}{($\downarrow$130)} & 190 \textcolor{green}{($\downarrow$47)} & 212 \textcolor{green}{($\downarrow$51)} \\
             ~~\textit{+TrimTail(70)}  & 6.40 / 5.37 & 160 \textcolor{green}{($\downarrow$80)} & 210 \textcolor{green}{($\downarrow$90)} & -30 \textcolor{green}{($\downarrow$240)} & 40 \textcolor{green}{($\downarrow$210)} & 112 \textcolor{green}{($\downarrow$125)} & 136 \textcolor{green}{($\downarrow$127)} \\
             ~~~\textit{+TrimTail(100)}  & 7.12 / 5.80 & 100 \textcolor{green}{($\downarrow$140)} & 140 \textcolor{green}{($\downarrow$160)} & -90 \textcolor{green}{($\downarrow$300)} & -20 \textcolor{green}{($\downarrow$270)} & 31 \textcolor{green}{($\downarrow$206)} & 59 \textcolor{green}{($\downarrow$204)} \\
        \midrule
        \multicolumn{8}{c}{Aishell (test) 1/4 subsample \& 160ms chunksize} \\
        \midrule
            Conformer (CTC) & \textbf{6.35} / \textbf{5.39} & 240 & 300 & 210 & 250 & 241 & 268 \\ 
             ~~\textit{+TrimTail(50)}  & 6.52 / 5.44 & 240 \textcolor{green}{($\sim$)} & 290 \textcolor{green}{($\downarrow$10)} & 40 \textcolor{green}{($\downarrow$170)} & 110 \textcolor{green}{($\downarrow$140)} & 192 \textcolor{green}{($\downarrow$49)} & 215 \textcolor{green}{($\downarrow$51)} \\
             ~~\textit{+TrimTail(70)}  & 6.84 / 5.62 & 180 \textcolor{green}{($\downarrow$60)} & 230 \textcolor{green}{($\downarrow$70)} & -30 \textcolor{green}{($\downarrow$240)} & 40 \textcolor{green}{($\downarrow$210)} & 119 \textcolor{green}{($\downarrow$122)} & 145 \textcolor{green}{($\downarrow$123)} \\
             ~~~\textit{+TrimTail(100)}  & 8.03 / 6.38 & 120 \textcolor{green}{($\downarrow$120)} & 180 \textcolor{green}{($\downarrow$120)} & -90 \textcolor{green}{($\downarrow$300)} & -20 \textcolor{green}{($\downarrow$270)} & 40 \textcolor{green}{($\downarrow$201)} & 71 \textcolor{green}{($\downarrow$197)} \\
        \midrule
        \multicolumn{8}{c}{Aishell (test) 1/8 subsample \& 640ms chunksize} \\
        \midrule
             Conformer (CTC) & 5.85 / \textbf{5.16} & 150 & 200 & 110 & 170 & 148 & 175 \\ 
             ~~\textit{+TrimTail(50)}  & \textbf{5.83} / \textbf{5.16} & 80 \textcolor{green}{($\downarrow$70)} & 130 \textcolor{green}{($\downarrow$70)} & -30 \textcolor{green}{($\downarrow$140)} & 40 \textcolor{green}{($\downarrow$130)} & 54 \textcolor{green}{($\downarrow$94)} & 80 \textcolor{green}{($\downarrow$95)} \\
             ~~\textit{+TrimTail(70)}  & 5.96 / 5.24 & 50 \textcolor{green}{($\downarrow$100)} & 110 \textcolor{green}{($\downarrow$90)} & -110 \textcolor{green}{($\downarrow$220)} & -30 \textcolor{green}{($\downarrow$200)} & -13 \textcolor{green}{($\downarrow$161)} & 16 \textcolor{green}{($\downarrow$159)} \\
             ~~~\textit{+TrimTail(100)}  & 6.54 / 5.52 & 20 \textcolor{green}{($\downarrow$130)} & 90 \textcolor{green}{($\downarrow$110)} & -180 \textcolor{green}{($\downarrow$290)} & -110 \textcolor{green}{($\downarrow$280)} & -93 \textcolor{green}{($\downarrow$241)} & -60 \textcolor{green}{($\downarrow$235)} \\
        \midrule
        \multicolumn{8}{c}{Librispeech (test\_clean) 1/4 subsample \& 640ms chunksize} \\
        \midrule
            Conformer (CTC) & 4.84 / 4.13 & - & - & - & - & - & - \\ 
             ~~\textit{+TrimTail(50)}  & \textbf{4.68} / \textbf{4.01} & - \textcolor{green}{($\downarrow$80)} & - \textcolor{green}{($\downarrow$40)} & - \textcolor{green}{($\downarrow$80)} & - \textcolor{green}{($\downarrow$40)} & - \textcolor{green}{($\downarrow$91)} & - \textcolor{green}{($\downarrow$77)} \\
             ~~\textit{+TrimTail(70)}  & 4.69 / 4.02 & - \textcolor{green}{($\downarrow$120)} & - \textcolor{green}{($\downarrow$80)} & - \textcolor{green}{($\downarrow$120)} & - \textcolor{green}{($\downarrow$80)} & - \textcolor{green}{($\downarrow$128)} & - \textcolor{green}{($\downarrow$113)} \\
             ~~~\textit{+TrimTail(100)}  & 4.82 / 4.12 & - \textcolor{green}{($\downarrow$160)} & - \textcolor{green}{($\downarrow$80)} & - \textcolor{green}{($\downarrow$120)} & - \textcolor{green}{($\downarrow$80)} & - \textcolor{green}{($\downarrow$149)} & - \textcolor{green}{($\downarrow$132)} \\
             ~~~\textit{+TrimTail(150)}  & 5.18 / 4.31 & - \textcolor{green}{($\downarrow$160)} & - \textcolor{green}{($\downarrow$120)} & - \textcolor{green}{($\downarrow$160)} & - \textcolor{green}{($\downarrow$80)} & - \textcolor{green}{($\downarrow$160)} & - \textcolor{green}{($\downarrow$142)} \\
             ~~~\textit{+TrimTail(200)}  & 5.24 / 4.38 & - \textcolor{green}{($\downarrow$160)} & - \textcolor{green}{($\downarrow$120)} & - \textcolor{green}{($\downarrow$160)} & - \textcolor{green}{($\downarrow$120)} & - \textcolor{green}{($\downarrow$174)} & - \textcolor{green}{($\downarrow$156)} \\
        \midrule
        \multicolumn{8}{c}{Librispeech (test\_other) 1/4 subsample \& 640ms chunksize} \\
        \midrule
            Conformer (CTC) & 11.77 / \textbf{10.62} & - & - & - & - & - & - \\ 
             ~~\textit{+TrimTail(50)}  & \textbf{11.68} / 10.73 & - \textcolor{green}{($\downarrow$80)} & - \textcolor{green}{($\downarrow$40)} & - \textcolor{green}{($\downarrow$80)} & - \textcolor{green}{($\downarrow$40)} & - \textcolor{green}{($\downarrow$87)} & - \textcolor{green}{($\downarrow$70)} \\
             ~~\textit{+TrimTail(70)}  & 11.73 / 10.68 & - \textcolor{green}{($\downarrow$120)} & - \textcolor{green}{($\downarrow$80)} & - \textcolor{green}{($\downarrow$120)} & - \textcolor{green}{($\downarrow$80)} & - \textcolor{green}{($\downarrow$124)} & - \textcolor{green}{($\downarrow$104)} \\
             ~~~\textit{+TrimTail(100)}  & 12.00 / 10.91 & - \textcolor{green}{($\downarrow$160)} & - \textcolor{green}{($\downarrow$80)} & - \textcolor{green}{($\downarrow$120)} & - \textcolor{green}{($\downarrow$80)} & - \textcolor{green}{($\downarrow$143)} & - \textcolor{green}{($\downarrow$123)} \\
             ~~~\textit{+TrimTail(150)}  & 12.68 / 11.34 & - \textcolor{green}{($\downarrow$160)} & - \textcolor{green}{($\downarrow$120)} & - \textcolor{green}{($\downarrow$160)} & - \textcolor{green}{($\downarrow$80)} & - \textcolor{green}{($\downarrow$155)} & - \textcolor{green}{($\downarrow$135)} \\
             ~~~\textit{+TrimTail(200)}  & 12.94 / 11.70 & - \textcolor{green}{($\downarrow$160)} & - \textcolor{green}{($\downarrow$120)} & - \textcolor{green}{($\downarrow$160)} & - \textcolor{green}{($\downarrow$120)} & - \textcolor{green}{($\downarrow$167)} & - \textcolor{green}{($\downarrow$144)} \\
             
        \bottomrule[1.5pt]
        \end{tabular}
    }
    \vspace{-11pt}
    \label{tab:Tmax}
\end{table*}

To evaluate the proposed \textit{TrimTail}, we carry out our experiments on the open-source Chinese Mandarin speech corpus Aishell-1~\cite{cocosda17/aishell-1} and English speech corpus Librispeech~\cite{ic15/librispeech}. We use WeNet~\cite{DBLP:journals/corr/abs-2203-15455}, an end-to-end speech recognition toolkit for all our experiments.


\textit{TrimTail} can be applied to any ASR model trained with any loss on any dataset without any extra effort. To demonstrate the effectiveness of our proposed method, we apply \textit{TrimTail} on a wide range of ASR models, losses, and datasets.
For each of our experiments, we keep the exact same training and testing settings as in open-sourced Aishell-1 and Librispeech recipes~\textsuperscript{\ref{web}}, including loss weight, model size, model regularization (weight decay, etc.), optimizer, learning rate schedule, data augmentation, etc.

\vspace{-10pt}
\subsection{Results and Discussions}
\label{sec:results}

In this section, we first report our results on Aishell-1 and Librispeech with various network structures, training objectives, and hyper-parameter $T_{max}$. We next perform controlled experiments to explore why \textit{TrimTail} tends to emit tokens faster and finally we show that apart from reducing token emission latency, one can also benefit from \textit{TrimTail} to achieve a much lower User Sensitive Delay.

\vspace{-10pt}
\subsubsection{TrimTail results on Aishell-1 and Librispeech}
\vspace{-5pt}

We first present the main results of \textit{TrimTail} in Table~\ref{tab:Tmax}. From Aishell-1 results, we find that \textit{TrimTail} significantly reduces Last Token emission Delay by 100ms (Transformer) $\sim$ 200ms (Conformer, either trained by CTC loss, Transducer loss, or both of them) with equal or even better WER. It is noteworthy that streaming ASR models that capture stronger contexts can emit the full hypothesis even before they were spoken, leading to a negative LTD. We also find larger $T_{max}$ leads to lower delay for all kinds of ASR models and losses. But when the $T_{max}$ is larger than a certain threshold, the WER starts to degrade due to the regularization being too strong. Overall, $T_{max}$ offers the flexibility of WER-latency trade-offs and this can be double-checked in Libripseech results. Besides, we also find \textit{TrimTail} even improves the recognition accuracy by 0.1 $\sim$ 0.2 on LibriSpeech while achieving roughly 100ms forward shift on token emission delay, demonstrating that \textit{TrimTail} does generalize well to any dataset without any extra effort.


\vspace{-10pt}
\subsubsection{Controlled Experiment}
\begin{figure}
    \centering
    \includegraphics[scale=0.35]{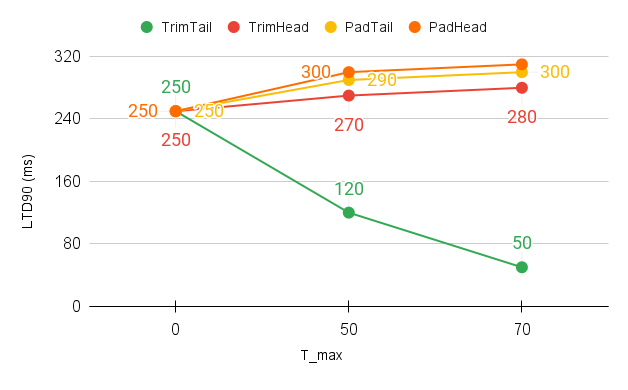}
    \vspace{-10pt}
    \caption{Controlled experiment on different length penalty strategies (Aishell-1).}
    \label{fig:ablation}
\end{figure}

\vspace{-4pt}
\label{sec:expablation}
To validate our conjectures that trimming trailing frames helps to \textbf{``squeeze''} alignment space and encourage the model to predict trailing tokens even before they were spoken, which further pushes forward the emission of all previous tokens due to the \textbf{monotonicity} of speech-text alignment. We perform three variants described in Fig.~\ref{fig:spectrim}(c)-(e) on Conformer (CTC). As shown in Fig.~\ref{fig:ablation}, Although TrimHead also decreases the alignment space, it suffers from speech-text mismatch and zero context when predicting leading tokens thus they were heavily delayed to help the model reduce the confusion of mismatched data and gather more context information. As for PadHead and PadTail, the increasing delay double-checked our arguments that squeezing alignment space helps to reduce the latency.

\vspace{-10pt}
\subsubsection{Algorithmic Improvement of USD}
\label{sec:improve_usd}
Finally, to theoretically analyze the maximum algorithmic improvement of User Sensitive Delay (AI-USD), we propose to cut fixed-length trailing speech at the inference stage, in which the maximum cutting length without significant WER degradation (i.e., more than 0.2) could be seen as the upper bound of AI-USD.

As shown in Table~\ref{tab:vad}, we obvious a WER degradation from 5.81 / 5.05 to 10.33 / 8.12 on baseline Conformer when the cutting length increased from 0ms to 400ms. On the contrary, model trained with \textit{TrimTail(50)} does not suffer from a 400ms tail missing. In other words, the AI-USD of \textit{TrimTail(50)} on Conformer (CTC) could be 400ms, considering the 130ms relative improvement on LTD90 (see Table~\ref{tab:Tmax}), the remaining 270ms improvement would result from the reduced endpoint delay. It is also noteworthy that when the cutting length comes to 500ms, the WER of baseline model has doubled while our method still maintains reasonable WER, indicating that \textit{TrimTail} does help to encourage the model to predict the trailing tokens even before they were spoken, which leads to a significant reduction of User Sensitive Delay.

\begin{table}[!htp]
    \centering
    \caption{WER comparison of different cutting lengths (Aishell-1).}
    \scalebox{0.83}{
        \begin{tabular}{ccc}
        \toprule[1.5pt]
            Cutting Length (ms) & Baseline & Baseline (\textit{+TrimTail(50)})\\
        \midrule
            0 & \textbf{5.81} / \textbf{5.05} & 5.83 / 5.09 \\
            200 & 5.93 / 5.16  & \textbf{5.83} / \textbf{5.09} \\
            300 & 7.26 / 5.89   & \textbf{5.85} / \textbf{5.09} \\
            400 & 10.33 / 8.12 & \textbf{6.0} / \textbf{5.16} \\
            500 & 13.08 / 10.95  & \textbf{7.79} / \textbf{6.03} \\
        \bottomrule[1.5pt]
    \end{tabular}
    }
    \label{tab:vad}
\end{table}



\vfill\pagebreak

\bibliographystyle{IEEEbib}
\bibliography{strings,refs}

\end{document}